\documentclass[conference,10pt]{IEEEtran}
\usepackage{amssymb,amsfonts,amsthm,amscd,dsfont,mathrsfs,verbatim}
\usepackage[cmex10]{amsmath}
\usepackage[tight,footnotesize]{subfigure}
\usepackage{graphicx}
\DeclareMathAlphabet{\mathpzc}{OT1}{pzc}{m}{it}

\newtheorem{propo}{Proposition}[section]
\newtheorem{lemma}[propo]{Lemma}

\newtheorem{coro}[propo]{Corollary}
\newtheorem{thm}[propo]{Theorem}

\def\sX{\mathbb{X}}
\def\sD{\mathbb{D}}
\def\hsD{\widehat{\mathbb{D}}}
\def\sZ{\mathbb{Z}}

\def\prob{{\mathbb P}}

\def\tY{\widetilde{Y}}

\def\st{\textup{stat. erg.}}

\def\St{\mathcal{S}}
\def\E{\mathbb{E}}
\def\eps{{\epsilon}}
\def\ve{{\varepsilon}}
\def\naturals{{\mathbb N}}
\def\integers{{\mathbb Z}}
\def\tW{\widetilde{W}}
\def\tK{K_1}
\def\hW{{\widehat{W}}}

\def\hy{{\widehat{y}}}

\begin{document}

\title{On the deletion channel with small deletion probability}

\author{\IEEEauthorblockN{Yashodhan Kanoria\IEEEauthorrefmark{1}\;\;\;
 and\;\;\;  Andrea Montanari\IEEEauthorrefmark{1}\IEEEauthorrefmark{2}}
\IEEEauthorblockA{Departments of Electrical Engineering\IEEEauthorrefmark{1}
 and Statistics\IEEEauthorrefmark{2}, Stanford University}
\IEEEauthorblockA{Email: \{ykanoria, montanari\}@stanford.edu}
}

\maketitle
\begin{abstract}
The deletion channel is the simplest point-to-point communication
channel that models lack of synchronization. Despite
significant effort, little is known about its capacity,
and even less about optimal coding schemes. In this paper we initiate
a new systematic approach to this problem, by demonstrating that
capacity can be computed in a series expansion for small deletion probability.
We compute two leading terms of this expansion, and
show that capacity
is achieved, up to this order, by i.i.d. uniform random distribution of
the input.

We think that this strategy can be useful in a number of capacity
calculations.
\end{abstract}

%
%

\section{Introduction}
\label{sec:intro}

The (binary) deletion channel accepts bits as inputs,
and deletes each transmitted bit independently with  probability $d$.
Computing or providing systematic approximations to its capacity is
one of the outstanding problems in information theory
\cite{MitzenmacherReview}. An important
motivation comes from the need to understand synchronization errors and
optimal ways to cope with them.

In this paper we suggest a new approach. We demonstrate that capacity
can be computed in a series expansion for small deletion probability,
by computing the first two orders of such an expansion. Our main result
is the following.
\begin{thm}\label{thm:main_theorem}
Let $C(d)$ be the capacity of the deletion channel with deletion
probability $d$. Then, for small $d$ and
any $\eps>0$,
\begin{align}
C(d)=1+ d\log d - A_1\, d+ O(d^{3/2-\epsilon}) \, ,\label{eq:MainFormula}
\end{align}
where $A_1\equiv \log(2e)-\sum_{l=1}^\infty 2^{-l-1}l\log l$.
Further, the iid Bernoulli$(1/2)$ process achieves capacity up to
corrections of order $O(d^{3/2-\epsilon})$.
\end{thm}
Logarithms here (and in the rest of the paper) are understood to be
in base $2$.
The constant $A_1$ can be easily evaluated to yield
$A_1\approx 1.154163765$.
While one might be skeptical about the concrete meaning
of asymptotic expansions of the type (\ref{eq:MainFormula}),
they often prove surprisingly accurate. For instance at $10\%$ deletion
probability,
Eq.~(\ref{eq:MainFormula}) is off the best lower bound
proved in \cite{Drinea07} by about $0.010$ bits.
More importantly they provide useful design insight. For instance, the above
result shows that Bernoulli$(1/2)$ is an excellent starting point for
the optimal input distribution. Next terms in expansion indicate
how to systematically modify the input distribution for $d>0$
\cite{InProgress}.
\begin{figure}
\centering
\includegraphics[width=3.in]{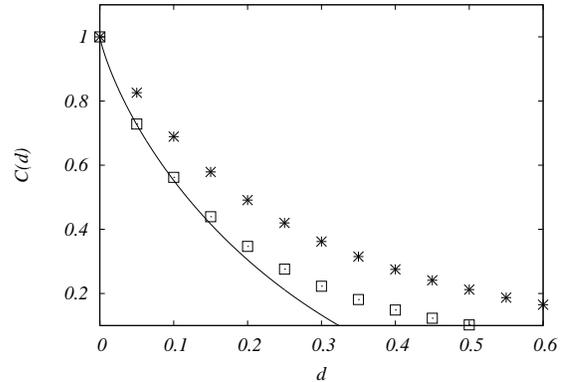}
\caption{Comparison of the asymptotic formula (\ref{eq:MainFormula})
(continuous line) with upper bounds from \cite{Fertonani09}
(stars $\ast$) and lower bounds from \cite{Drinea07} (squares, $\Box$). The $O(d^{3/2-\eps})$ term in (\ref{eq:MainFormula}) was simply dropped.}\label{fig_bd}
\vspace{-0.3cm}
\end{figure}

We think the strategy adopted here might be useful in other
information theory problems. The underlying philosophy is
that whenever capacity is known for a specific value of
the channel parameter, and the corresponding
optimal input distribution is unique and well characterized,
it should be possible to compute an asymptotic expansion around that value.
Here the special channel is the perfect channel,
i.e. the deletion channel with deletion probability $d=0$. The
corresponding input distribution is the iid Bernoulli$(1/2)$ process.
%
%
\subsection{Related work}

Dobrushin \cite{Dobrushin} proved a coding theorem for the deletion channel,
and other channels with synchronization errors. He showed that the maximum
rate of reliable communication is given by the  maximal mutual information per
bit, and proved that this can be achieved through a random coding scheme.
This characterization has so far found limited use in proving concrete
estimates.
An important exception is provided by the work of Kirsch and Drinea
\cite{KirschDrinea} who use Dobrushin coding theorem to prove
lower bounds on the capacity of channels with deletions and
duplications.
We will also use Dobrushin theorem in a crucial way, although
most of our effort will be devoted to proving upper bounds on the capacity.

Several capacity bounds have been developed over the last few years,
following alternative approaches, and are surveyed in
\cite{MitzenmacherReview}.
In particular, it has been proved that $C(d)=\Theta(1-d)$
as $d\to 1$. However determining the asymptotic behavior in
this limit (i.e. finding a constant $B_1$ such that
$C(d) = B_1(1-d)+o(1-d)$) is an open problem.
When applied to the small $d$ regime, none of the known upper bounds
actually captures the correct behavior (\ref{eq:MainFormula}).
As we show in the present paper, this behavior can be
controlled exactly.

When this paper was nearing submission, a preprint
by Kalai, Mitzenmacher and Sudan \cite{KMS} was posted online,
proving a statement analogous to Theorem \ref{thm:main_theorem}.
The result of \cite{KMS} is however not the same as in Theorem
\ref{thm:main_theorem}: only the $d\log d$ term of the series
is proved in \cite{KMS}. Further,
the two proofs are based on very different approaches.
%
%
\section{Preliminaries}

For the reader's convenience, we restate here some known results
that we will use extensively, along with with some definitions
and auxiliary lemmas.

Consider a sequence of channels $\{W_n\}_{n\ge 1}$, where $W_n$ allows exactly
$n$ inputs bits, and deletes each bit independently with probability $d$.
The output of $W_n$ for input $X^n$ is a binary vector denoted by $Y(X^n)$.
The length of $Y(X^n)$ is a binomial random variable.
We want to find maximum rate at which we can send information
over this sequence of channels
with vanishingly small error probability.

The following characterization follows from \cite{Dobrushin}.
\begin{thm}\label{lemma:cap_limit}
Let
\begin{align}
C_n= \frac{1}{n}\max_{p_{X^n}}\, I(X^n; Y(X^n)) \, .
\end{align}
Then, the following limit exists
\begin{align}
C=\lim_{n\rightarrow \infty}C_n = \inf_{n\ge 1} C_n\, ,
\label{eq:cap_defined}
\end{align}
and is equal to the capacity of the deletion channel.
\end{thm}
\begin{IEEEproof}
This is just a reformulation of Theorem 1 in \cite{Dobrushin},
to which we  add the remark $C = \inf_{n\ge 1} C_n$, which is of independent
interest.
In order to prove this fact, consider the channel $W_{m+n}$,
and let $X^{m+n}= (X_1^m,X_{m+1}^{m+n})$ be its input.
The channel $W_{m+n}$ can be realized as follows.
First the input is passed through a channel
$\tW_{m+n}$ that introduces deletions independently in the two strings
$X_{1}^m$ and $X_{m+1}^{m+n}$ and outputs
$\tY(X_1^{m+n})\equiv (Y(X_{1}^m),|,Y(X_{m+1}^{m+n}))$
where $|$ is a marker. Then the marker is removed.

This construction proves that $W_{m+n}$ is  physically degraded
with respect to $\tW_{m+n}$, whence
\begin{eqnarray*}
(m+n)C_{m+n}&\le &\max_{p_{X^{m+n}}} I(X^{m+n};\tY(X_{1}^{m+n}))\\
&\le & mC_m+nC_n\, .
\end{eqnarray*}
Here the last inequality follows from the fact that $\tW_{m+n}$ is the product
of two independent channels, and hence the mutual information is maximized
by a product input distribution.

Therefore the sequence $\{nC_n\}_{n\ge 1}$ is sub-additive, and
the claim follows
from Fekete's lemma.
\end{IEEEproof}

A last useful remark is that, in computing capacity, we can
assume $(X_1,\dots,X_n)$ to be $n$ consecutive coordinates of a stationary
ergodic process.
\begin{lemma}\label{lemma:stationary_suffices}
Let $\sX= \{X_i\}_{i\in\integers}$ be a stationary and ergodic
process, with $X_i$ taking values in $\{0,1\}$.  Then the limit
$I(\sX)=\lim_{n \rightarrow \infty} \frac{1}{n}I(X^n; Y(X^n))$ exists and
\begin{align}
C=\max_{\sX\; \st} I(\sX)\,.
\end{align}
\end{lemma}
\begin{IEEEproof}
Take any stationary $\sX$, and let $I_n= I(X^n; Y(X^n))$. Notice that
 $Y(X_1^n)-X_1^n-X_{n+1}^{n+m}-Y(X_{n+1}^{n+m})$ form a Markov chain.
Define $\tY(X^{n+m})$ as in the proof of Theorem \ref{lemma:cap_limit}.
As before we have $I_{n+m} \leq I(X^{n+m},\tY(X^{n+m})) \leq I(X_1^m;\tY(X_1^m)) +I(X_{m+1}^{m+n};Y(X_{m+1}^{m+n})) = I_m+I_n$.
(the last identity follows by
stationarity of $\sX$). Thus $I_{m+n}\leq I_n+I_m$ and the limit
$\lim_{n\to\infty}I_n/n$
exists by Fekete's lemma, and is equal to $\inf_{n\geq 1}I_n/n$.

Clearly, $I_{n} \leq C_n$ for all $n$.
Fix any $\ve>0$. We will construct a process $\sX$ such that
\begin{align}
 I_{N}/N\geq  C - \ve \qquad \forall \; N>N_0(\ve)\, ,
\label{eq:liminf_closeto_limsup}
\end{align}
thus proving our claim.

Fix $n$ such that
$C_n \geq  C - \ve/2$. Construct $\sX$ with
iid blocks of length $n$ with common distribution $p^*(n)$ that
achieves the supremum in the definition of $C_n$.
In order to make this process stationary,
we make the first complete block to the
right of the position $0$ start at position
$s$ uniformly random in $\{1,2,\dots,n\}$.
We call the position $s$ the offset.
The resulting process is clearly stationary and ergodic.

Now consider $N=kn+r$ for some $k \in \naturals$
and $r \in \{0, 1, \ldots, n-1\}$. The vector $X_1^N$ contains
at least $k-1$ complete blocks of size $n$, call them $X(1), X(2),
\ldots, X(k-1)$ with $X(i) \sim p^*(n)$. The block $X(1)$
starts at position $s$. There will be  further $r+n - s+1$ bits
at the end, so that $X_1^N=(X_1^{s-1}, X(1), X(2), \ldots, X(k-1),
X_{s+kn}^N)$.
Abusing notation, we write $Y(i)$ for $Y(X(i))$.
Given the output $Y$, we define
$\tY= ( Y(X_1^{s-1})| Y(1) | Y(2) | \ldots |Y(k-1)|Y(X_{s+(k-1)n}^N))$,
by introducing $k$ synchronization symbols $|$.
There are at most $(n+1)^k$ possibilities for $\tY$ given $Y$
(corresponding to potential placements of synchronization symbols).
Therefore we have
\begin{align*}
H(Y) &= H(\tY) - H(\tY|Y)\\
&\geq H(\tY) - \log((n+1)^k)\\
&\geq (k-1)H(Y(1)) - k\log(n+1)\, ,
\end{align*}
where we used the fact that the $(X(i),Y(i))$'s are iid.
Further
\begin{align*}
H(Y|X^N) \leq H(\tY|X^N) \leq (k-1)H(Y(1)|X(1)) + 2n\, ,
\end{align*}
where the last term accounts for bits outside the blocks.
We conclude that
\begin{align*}
I(X^N;Y(X^N)) &= H(Y) - H(Y|X^N)\\
&\geq (k-1)nC_n - k \log(n+1) - 2n\\
&\geq N(C_n - \ve/2) \, ,
\end{align*}
provided $\log(n+1) /n < \ve / 10 $, $N > N_0\equiv 10n/\ve$.
Since $C_n\ge C-\ve/2$, this in turn implies
Eq.~(\ref{eq:liminf_closeto_limsup}).
\end{IEEEproof}

%
%

\section{Proof of the main theorem: Outline}
\label{sec:outline}

In this section we provide the proof of Theorem \ref{thm:main_theorem}.
We defer the proof of several technical lemmas to the next section.

The first step consists in proving achievability by
estimating $I(\sX)$ for the iid Bernoulli$(1/2)$ process.
\begin{lemma}\label{lemma:iidhalf}
Let $\sX^*$ be the iid {\rm Bernoulli}$(1/2)$ process.
For any $\eps >0$, we have
\begin{align}
I(\sX^*) = 1+ d\log d - A_1\, d + O(d^{2-\eps})\, .
\end{align}
\end{lemma}
Lemma \ref{lemma:stationary_suffices} allows us to restrict our
attention to stationary ergodic processes in proving the converse.
In light of Lemma \ref{lemma:iidhalf}, we can further restrict
consideration to processes $\sX$ satisfying $I(\sX) > 1+2d\log d$
and hence $H(\sX) > 1+2d\log d$ (here and below, for a process
$\sX$, we denote by $H(\sX)$ its \emph{entropy rate}).

Given a (possibly infinite) binary sequence,
a \emph{run} of $0$'s (of $1$'s) is a maximal subsequence of consecutive $0$'s
($1$'s), i.e. an subsequence of $0$'s bordered by $1$'s
(respectively, of $1$'s bordered by $0$'s).
Denote by $\St$ the set of all stationary ergodic processes and
by $\St_L$ the set of stationary ergodic processes such
that, with probability one, no run has length larger than $L$.
The next lemma shows that we don't lose much by restricting
ourselves to $\St_{L^*}$ for large enough $L^*$.
\begin{lemma}\label{lemma:small_loss_by_restricting_runs}
For any $\eps>0$ there exists  $d_0=d_0(\eps)>0$
such that the following happens for all $d < d_0$.
For any $\sX \in \St$ such that $H(\sX) > 1 + 2d \log d$ and for any
$L^* > \log (1/d)$, there exists
$\sX_{L^*} \in \St_{L^*}$  such that
\begin{align}
I(\sX) \leq I(\sX_{L^*}) + d^{1/2-\epsilon}(L^*)^{-1}\log L^*\, .
\end{align}
\end{lemma}
We are left with the problem of bounding $I(\sX)$ from above for all
$\sX \in \St_{L^*}$. The next lemma establishes such a bound.
\begin{lemma}\label{lemma:converse_for_restricted_runs}
For any $\eps>0$ there exists  $d_0=d_0(\eps)>0$ such that the following
happens.
For any $L^* \in \mathbb{N}$ and any $\sX \in \St_{L^*}$
if $d < d_0(\epsilon)$, then
\begin{align}
I(\sX) \leq 1+ d\log d - A_1d + d^{2-\epsilon}(1+d^{1/2}L^*)\, .
\end{align}
\end{lemma}
\begin{IEEEproof}[Proof of Theorem \ref{thm:main_theorem}]
Lemma \ref{lemma:iidhalf} shows achievability.
The converse follows from Lemmas \ref{lemma:small_loss_by_restricting_runs}
and \ref{lemma:converse_for_restricted_runs} with
$L^*= \lfloor 1/d \rfloor$.
\end{IEEEproof}
%
%
\section{Proofs of the Lemmas}

In Section \ref{subsec:run_charac} we characterize any stationary
ergodic $\sX$ in terms of its `bit perspective' and `block perspective'
run-length distributions, and show that these distributions must be
close to the distributions obtained for the iid Bernoulli$(1/2)$ process.
In Section \ref{subsec:modified_deletion} we construct a modified
deletion process that allows accurate estimation of $H(Y|X^n)$ in
the small $d$ limit. Finally, in Section \ref{subsec:lemma_proofs}
we present proofs of the Lemmas quoted in Section \ref{sec:outline}
using the tools developed.

We will often write $X_{a}^b$ for the random vector
$(X_a,X_{a+1},\dots, X_b)$ where the $X_i$'s are distributed according
to the process $\sX$.
%
%
\subsection{Characterization in terms of runs}
\label{subsec:run_charac}

Consider a stationary ergodic process $\sX$.
Without loss of generality we can assume that almost surely all runs
have finite length (by ergodicity and stationarity this only excludes the
constant $0$ and constant $1$ processes).
Let $L_0$ be the length of the
run containing position $0$ in $\sX$.
 Let $L_1$ be the length of
first run to occur to the right of position $0$ in $\sX$ and,
in general, let $L_i$ be the length of the $i$-th run to the right of
position $0$.  Let $p_{L,\sX}$
denote the limit of the empirical distribution of $L_1, L_2, \ldots,L_K$,
as $K\to\infty$. By ergodicity
$p_{L,\sX}$ is a well defined probability
distribution on $\naturals$.
We call $p_{L,\sX}$ the \emph{block-perspective} run length distribution
for obvious reasons, and use $L$ to denote a random variable
drawn according to $p_{L,\sX}$.

It is not hard to see that,
for any $l\ge 1$,
\begin{align}
\prob(L_0= l)=\frac{lp_{L,\sX}(l)}{\E[L]} \; .
\end{align}
In other words $L_0$ is distributed according to the size biased
version of $p_{L,\sX}$.
We call this the \emph{bit perspective} run  length distribution,
and shall often drop the subscript $\sX$ when clear from the context.
Notice that since $L_0$ is a well defined and almost surely finite,
we have $\E[L] < \infty$. It follows that the empirical distribution of
run lengths in
$X_1^n$ also converges to $p_{L,\sX}$ almost surely, since
the first and last run do not matter in the limit.

If $L_0^+,L_1,\dots, L_K$ are the run lengths in the block
$X_0^n$, it is clear that $H(X_0^n) \le 1+H(L_1,\dots,L_{K_n},K_n)$
(where one bit is needed to remove the $0,1$ ambiguity).
By ergodicity $K_n/n\to 1/\E[L]$ almost surely as $n\to\infty$.
This also implies $H(K_n)/n\to 0$.
Further, 
$\limsup_{n\rightarrow \infty} H(L_1,\dots,L_{K_n})/n \le \lim_{n\rightarrow \infty}
H(L) K_n/n = H(L)/\E[L] $.
If $H(\sX)$ is the entropy rate of the process $\sX$,
by taking the $n\to\infty$ limit, it is easy to deduce that
\begin{align}
H(\sX) \leq \frac{ H(L) }{\E[L]} \, ,
\label{eq:run_hx_upper_bd}
\end{align}
with equality if and only if $\sX$ consists of iid runs with common
distribution $p_L$.

For convenience of notation, define $\mu(\sX) \equiv \E[L]$.
We know that given $\E[L]=\mu$, the probability distribution
with largest possible entropy $H(L)$ is geometric with mean $\mu$, i.e.
$p_L(l) = (1-1/\mu)^{l-1}1/\mu $ for all $ l \geq 1$, leading to
\begin{align}
\frac{H(L)}{\E[L]} \leq  -\big(1 - \frac{1}{\mu}\big) \log
\big(1 - \frac{1}{\mu}\big) -
\frac{1}{\mu}\log \frac{1}{\mu} \equiv h(1/\mu) \, .
\label{eq:BoundMu}
\end{align}
Here we introduced the notation $h(p) = -p \log p - (1-p) \log(1-p)$
for the binary entropy function.

In light of Lemma \ref{lemma:iidhalf}  we can
restrict ourselves to $H(\sX) > 1+2\, d\log d$.
Using this, we are able to obtain sharp bounds on $p_L$ and
$\mu(\sX)$.
\begin{lemma}
There exists $d_0>0$
such that, for  any $\sX\in\St$ with $H(\sX) > 1 + 2d \log d$,
\begin{align}
|\mu(\sX) - 2 | \leq \sqrt{100 \; d \log (1/d)} \, .
\end{align}
for all $d < d_0$.
\label{lemma:mean_closeto2}
\end{lemma}
\begin{IEEEproof}
By Eqs. (\ref{eq:run_hx_upper_bd}) and (\ref{eq:BoundMu}),
we have $h(1/\mu)\ge 1+2 d\log d$. By  Pinsker's inequality
$h(p)\le 1-(1-2p)^2/(2\ln 2)$, and therefore
$|1-(2/\mu)|^2\le (4\ln 2)d\log(1/ d)$.
The claim follows from simple calculus.
\end{IEEEproof}

\begin{lemma}\label{lemma:L_TV}
There exists  $K' < \infty$ and $d_0 >0$ such that, for any
$\sX\in\St$ with $H(\sX) > 1 + 2d \log d$, and any $d<d_0$,
\begin{align}
\sum_{l=1}^\infty \left|p_L(l) - \frac{1}{2^l} \right| \leq
K'\sqrt{ d \log (1/d)}\, .
\end{align}
\end{lemma}
\begin{IEEEproof}
Let $p_L^*(l) = 1/2^l, \ l \geq 1$ and recall that
$\mu(\sX) =\E[L]=\sum_{l\ge 1}
p_L(l)l$. An explicit calculation yields
\begin{align}
H(p_L)= \mu(\sX)- D(p_L || p_L^*) \, .
\label{eq:Lentropy}
\end{align}
Now, by Pinsker's inequality,
\begin{align}
D(p_L || p_L^*) \geq \frac{2}{\ln 2}||p_L-p_L^*||_{\rm TV}^2 \, .
\label{eq:pinsker}
\end{align}
Combining Lemma \ref{lemma:mean_closeto2},
and Eqs.~(\ref{eq:run_hx_upper_bd}), (\ref{eq:Lentropy})
and (\ref{eq:pinsker}), we
get the desired result.
\end{IEEEproof}

\begin{lemma}\label{lemma:L0_TV}
There exists  $K'' < \infty$ and $d_0 >0$ such that, for any
$\sX\in\St$ with $H(\sX) > 1 + 2d \log d$,  and any $d<d_0$,
\begin{align}
\sum_{l=1}^\infty\left|\prob(L_0=l) - \frac{l}{2^{l+1}} \right| \leq K''\sqrt{ d (\log (1/d))^3}\, .
\end{align}
\end{lemma}
\begin{IEEEproof}
Let $l_0 = \lfloor -\log(K'\sqrt{ d \log (1/d)}) \rfloor$. It follows from Lemma \ref{lemma:L_TV}
that
\begin{align}
\sum_{l=1}^{l_0} \left|p_L(l) - \frac{1}{2^l} \right| \leq
K'\sqrt{ d \log (1/d)}\, ,
\end{align}
which in turn implies
\begin{align}
\sum_{l=0}^{l_0} l p_L(l) \geq \sum_{l=0}^{l_0-1} \frac{l}{2^l}\, .
\label{eq:bound_firstl0_lpl}
\end{align}
Summing the geometric series, we find that there exists a constant
$\tK<\infty$ such that
\begin{align}
 \sum_{l=l_0}^{\infty} \frac{l}{2^l} = (l_0+1) 2^{1-l_0} \le
\tK \sqrt{d (\log (1/ d))^3}\, .
\label{eq:bound_afterl0_lby2tol}
\end{align}

Using the identity $\sum_{l=0}^{\infty} l\, 2^{-l}=2$, together
with Eqs.~(\ref{eq:bound_firstl0_lpl}) and (\ref{eq:bound_afterl0_lby2tol}),
we get
\begin{align}
\sum_{l=0}^{l_0} l p_L(l) \geq 2-\tK \sqrt{d (\log (1/d))^3}\, .
\end{align}
Combining this result with
Lemma \ref{lemma:mean_closeto2}, we conclude (eventually enlarging the constant
$\tK$)
\begin{align}
\sum_{l=l_0+1}^{\infty} l p_L(l) \leq 2\tK \sqrt{d (\log (1/d))^3}\, .
\end{align}
Using this result together with
Eq.~(\ref{eq:bound_afterl0_lby2tol}), we get
\begin{align}
\sum_{l=l_0+1}^{\infty}| l p_L(l) -\frac{l}{2^l}| \leq 4\tK \sqrt{d (\log (1/d))^3}\, .
\label{eq:nonnormalized_L0_TV_afterl0}
\end{align}

From a direct application of Lemma \ref{lemma:L_TV} it follows that
there exists a constant $K_2<\infty$, such that
\begin{align}
\sum_{l=1}^{l_0}\Big| l p_L(l) -\frac{l}{2^l}
\Big| \leq K_2 \sqrt{d (\log (1/d))^3}\, .
\label{eq:nonnormalized_L0_TV_beforel0}
\end{align}
and therefore summing Eqs.  (\ref{eq:nonnormalized_L0_TV_beforel0}) and
(\ref{eq:nonnormalized_L0_TV_afterl0})
\begin{align}
\sum_{l=1}^{\infty}\Big| \frac{l p_L(l)}{2} -\frac{l}{2^{l+1}}\Big| \leq
2(K_1+K_2) \sqrt{d (\log (1/d))^3}\, .
\label{eq:nonnormalized}
\end{align}
We know that $\prob(L_0=l) = lp_L(l) / \mu(\sX)$. The proof is completed by
using Eq.~(\ref{eq:nonnormalized}) and bounding $\mu(\sX)$ with the
Lemma \ref{lemma:mean_closeto2}.
\end{IEEEproof}
%
%
\subsection{A modified deletion process}
\label{subsec:modified_deletion}

We define an auxiliary sequence of channels
$\widehat{W}_n$ whose output --denoted by $\widehat{Y}(X^n)$--
is obtained by modifying the deletion channel output in the following way.
If an `extended run' (i.e. a run $\mathcal{R}$ along with one additional bit at each
end of $\mathcal{R}$) undergoes more than one deletion under the deletion
channel, then $\mathcal{R}$ will experience no deletion in channel $\widehat{W}_n$, i.e. the corresponding bits are
\textit{present} in $\widehat{Y}(X^n)$. Note that (deletions in) the additional bits at the ends are not affected.

Formally, we construct this sequence of channels as follows
when the input is a stationary process $\sX$.
Let $\sD$ be an iid Bernoulli$(d)$ process, independent
of $\sX$, with $D_1^n$ being the $n$-bit vector that contains a $1$ if
and only if the corresponding bit in $X^n$ is deleted by the channel $W_n$.
We define $\widehat{\sD}(\sD, \sX)$ to be the process containing
a subset of the $1$s in $\sD$. The process $\hsD$ is obtained
by deterministically flipping some of the $1$s in $\sD$ as
described above, simultaneously for all runs. The output of the channel $\hW_n$ is simply defined by
deleting from $X^n$ those bits whose positions correspond to $1$s in
$\hsD$.

Notice that $(\sX,\sD,\hsD)$ are jointly stationary.
The sequence of channels $W_n$ are defined by $\sD$, and
the coupled sequence of channels $\widehat{W}_n$ are defined
by $\hsD$. We emphasize that $\widehat{\sD}$ is a function of
$(\sX,\sD)$.
Let $\sZ \equiv \sD \oplus \widehat{\sD}$ (where $\oplus$ is componentwise
sum modulo $2$). The process $\sZ$ is  stationary
with $\prob(Z_0=1)\equiv z=\E[d- d(1-d)^{L_0+1}] \leq 2\,d^2\, \E[L_0]$.
Note that $z=O(d^2)$ for $\E[L_0]=O(1)$.

The following lemma shows the utility of the modified deletion process.
\begin{lemma}
Consider any $\sX \in \St$ such that $\E[L_0 \log L_0] < \infty$. Then
\begin{align}
\lim_{n\to\infty}\frac{1}{n}H(\widehat{D}^n|X^n, \widehat{Y}^n)=
d\, \E[\log L_0] - \delta\, ,
\end{align}
where
$0\leq \delta = \delta (d, \sX)\leq 2d^2\E[L_0\log L_0]$.
\label{lemma:hatD_givenxy}
\end{lemma}
\begin{IEEEproof}
Fix a channel input $x^n$ and any
possible output $\hy =\widehat{y}(x^n)$
(i.e. an output that occurs with positive probability under $\widehat{W}_n$).
The proof consists in estimating (the logarithm of)
the number of realizations of $\widehat{D}^n$ that
might lead to the input/ouput pair $(x^n,\hy)$, and then taking the expectation
over $(x^n,\hy)$.

Proceeding from left to right, and using the constraint on $\hsD$,
we can map unambiguously each run in $\widehat{y}$ to
one or more runs in $x^n$, that gave rise to it through the deletion process.
Consider a run of length $\ell$ in $\hy$.
If there is a unique `parent' run, it must have length
$\ell$ or $\ell+1$.
If the length of the parent run is $\ell$, then no deletion
occurred in this run, and hence the contribution to
$H(\widehat{D}^n|x^n, \widehat{y})$ of such runs vanishes.
If the length of the parent run is $\ell+1$,
one bit was deleted by $\widehat{W}^n$ and each of the
$\ell+1$ possibilities is equally likely, leading to a
contribution $\log(\ell+1)$ to $H(\widehat{D}^n|x^n, \widehat{y})$.

Finally, if there are multiple parent runs of lengths $l_1, l_2, \ldots, l_k$,
they must be separated by single bits of taking the opposite
value in $x^n$, all of which were deleted.
It also must be the case that $\sum_{i=1}^kl_i = \ell$ i.e. there is no
ambiguity in $\widehat{D}^n$.
This also implies $l_1<\ell$.

Notice that the three cases described corresponds to three different lengths
for the run in $\hy$. This allows  us to sequentially
associate runs in $\widehat{y}$ with runs in $x^n$, as claimed.

By the above argument,
$H(\widehat{D}^n|x^n, \widehat{y}^n)
= \sum_{r\in {\cal D}} \log(\ell_r)$ where ${\cal D}$ is the set
of runs on which deletions did occur, and $\ell_r$ are their
lengths. Using the definition of $\hsD$, the sum can be expressed
as $\sum_{i=1}^n\widehat{D}_i\log(\ell_{(i)})$, with $\ell_{(i)}$
the length of the run containing the $i$-th bit.
Using the definition of $\hsD$, we get
$\prob(\widehat{D}_i=1)= d(1-d)^{\ell_{(i)}+1} \in (d-(\ell_{(i)}+1)d^2, d)$
(except for the last and first block in $x^n$, that can be disregarded).
Taking expectation and letting $n\to\infty$ we get
the claim.
\end{IEEEproof}
\begin{coro}
Under the assumptions of the last Lemma, and
denoting by  $h(p)$ the binary entropy function, we have
\begin{align*}
\lim_{n \rightarrow \infty} \frac{1}{n}H(Y(X^n)|X^n) = h(d) - d\,
\E[\log L_0] + \delta \, ,
\end{align*}
where  $-2h(z)\leq \delta=
\delta (d, \sX) \leq 2d^2\E[L_0\log L_0]+2h(z)$ and $z= d- \E[d(1-d)^{L_0+1}]$.
\label{coro:hygivenx}
\end{coro}
\begin{IEEEproof}
By definition, $D^n$ is independent of $X^n$.
We have, for $Y= Y(X^n)$,
\begin{align*}
H(Y|X^n)&= H(D^n|X^n) - H(D^n|X^n, Y)\\
&= nh(d) -  H(\widehat{D}^n|X^n, \widehat{Y}) + n \delta_1 \, ,
\end{align*}
with $|\delta_1(d, \sX) |\leq 2 H(Z^n)/n\le 2 h(z)$.
In the second equality we used the fact that the pairs $((X^n, Y, D^n), (X^n, \widehat{Y}, \widehat{D}^n))$ and
$((X^n, Y), (X^n, \widehat{Y}))$ are both of the form $(A, B)$ such that $A$ is a function of
$(B, Z^n)$ and $B$ is a function of $(A, Z^n)$, $\Rightarrow |H(A) - H(B)| \leq H(Z^n)$.
\end{IEEEproof}
%
%
\subsection{Proofs of Lemmas \ref{lemma:iidhalf},  \ref{lemma:small_loss_by_restricting_runs} and \ref{lemma:converse_for_restricted_runs}}
\label{subsec:lemma_proofs}

\begin{IEEEproof}[Proof of Lemma \ref{lemma:iidhalf}]
Clearly, $\sX^*$
has run length distribution $p_L(l) = 2^{-l}$, $l \geq 1$. Moreover,
$Y(X^{*,n})$ is also a iid Bernoulli$(1/2)$ string of length
$\sim \textup{Binomial}(n,1-d)$.
Hence, $H(Y)= n (1-d) +O(\log n)$. We now use the estimate of $H(Y|X^{*,n})$
from Corollary \ref{coro:hygivenx}.  We have $z= O(d^2)$ and $\E[L_0\log L_0] < \infty$, leading to
\begin{align*}
H(Y|X^{*,n})  = n( h(d) - d\, \E [ \log L_0] + O(d^{2-\eps}) )+o(n)\, .
\end{align*}
Computing $H(Y) - H(Y|X^{*,n})$, we get the claim.
\end{IEEEproof}

\begin{IEEEproof}[Proof of Lemma \ref{lemma:small_loss_by_restricting_runs}]
We construct $\sX_{L^*}$ by flipping a bit each time it is the $(L^*+1)$-th
consecutive bit with the same value (either $0$ or $1$).
The density of such bits  in $\sX$ is upper bounded by
$\alpha=\prob(L_0>L^*)/L^*$. The expected fraction of bits in the
channel output $Y_{L^*}=Y(X^n_{L^*})$ that have been flipped relative to
$Y=Y(X^n)$ (output of the same channel realization
with different input) is also
at most $\alpha$. Let $F=F(\sX, \sD)$ be the binary vector  having the same
length as $Y$, with a $1$ wherever the corresponding bit in $Y_{L^*}$
is flipped relative to $Y$, and $0$s elsewhere. The expected fraction of $1$'s
in $F$ is at most $\alpha$. Therefore
\begin{align}
H(F) \leq n (1-d) h(\alpha) + \log (n+1)\, .
\label{eq:hflips_bound}
\end{align}
Notice that $Y$ is a deterministic function of $(Y_{L^*}, F)$ and $Y_{L^*}$ is a deterministic function of $(Y, F)$, whence
\begin{align}
|H(Y) - H(Y_{L^*})| \leq H(F)\, .
\label{eq:hy_bound_flips}
\end{align}
Further, $\sX-\sX_{L^*}-X_{L^*}^n-Y_{L^*}$ form a Markov chain,
and $\sX_{L^*}$, $X_{L^*}^n$ are deterministic functions of $\sX$.
Hence, $H(Y_{L^*} | X_{L^*}^n) = H(Y_{L^*} | \sX)$. Similarly,
$H(Y | X^n) = H(Y| \sX)$. Therefore (the second step is analogous to
Eq.~(\ref{eq:hy_bound_flips}))
\begin{align}
|H(Y_{L^*} | X_{L^*}^n) & - H(Y | X^n)| =\label{eq:hygivenx_bound_flips}\\
&=\;
|H(Y_{L^*} | \sX) - H(Y | \sX)|
\leq \;  H(F)\, . \nonumber
\end{align}
It follows from Lemma \ref{lemma:L0_TV} and $L^* > \log(1/d)$ that
$\alpha \leq 2K'' \sqrt{d (\log (1/d))^3}/L^*$ for sufficiently small $d$.
Hence, $h(\alpha) \leq d^{1/2-\epsilon}  \log L^* /(2 L^*)$ for $d < d_0( \epsilon)$,
for some $d_0( \epsilon) > 0$.
The result follows by combining Eqs.~(\ref{eq:hflips_bound}),
(\ref{eq:hy_bound_flips}) and (\ref{eq:hygivenx_bound_flips}) to bound
$|I(\sX) - I(\sX_{L^*})|$.
\end{IEEEproof}

\begin{IEEEproof}[Proof of Lemma \ref{lemma:converse_for_restricted_runs}]
If $H(\sX) \leq 1+ 2d \log d$, we are done. Else we proceed as follows.
We know that $Y(X^n)$ contains Binomial$(n,1-d)$ bits, leading immediately to
\begin{align}
H(Y) \leq n(1-d) + \log(n+1) \, .
\end{align}
We use the lower bound on $H(Y|X^n)$ from Corollary \ref{coro:hygivenx}.
We have $z \leq 2 d^2 \E [ L_0]$. It follows from Lemma \ref{lemma:L0_TV}
that $\E [L_0] \leq K_1 (1 + \sqrt{d (\log (1/d))^3} L^*)$, leading to $h(z) \leq .5d^{2-\epsilon}(1+(1/2)d^{1/2}L^*)$ for all $d < d_0$, where $d_0=
d_0(\epsilon)>0$. Thus, we have the bound
\begin{align*}
\lim_{n \rightarrow \infty}\frac{1}{n} H(Y|X^n)  \geq h(d) - d\E [ \log L_0] -  d^{2-\epsilon}(1+.5d^{1/2}L^*)
\end{align*}
Using Lemma \ref{lemma:L0_TV}, we have $|\E [ \log L_0] - \sum_{l=1}^\infty 2^{-l-1}l\log l| = o(d^{(1/2)-\epsilon}\log L^*)$.
The result follows.
\end{IEEEproof}
\vskip8pt

{\bf Acknowledgments.}
Y. Kanoria is supported by
a 3Com Corporation Stanford Graduate Fellowship. Y. Kanoria and A. Montanari were
supported by NSF, grants
CCF-0743978 and CCF-0915145, and a Terman fellowship.

\bibliographystyle{IEEEtran}

\end{document}